\documentclass[12pt]{article}
\usepackage{amssymb}

\begin{document}

\title{Harmonic fields on the extended projective disc and a problem in optics}
\author{Thomas H. Otway\thanks{%
email: otway@ymail.yu.edu} \\
\\
\textit{Department of Mathematics, Yeshiva University,}\\
\textit{\ \ New York, New York 10033}}
\date{}
\maketitle

\begin{abstract}
The Hodge equations for 1-forms are studied on Beltrami's projective
disc model for hyperbolic space. Ideal points lying beyond
projective infinity arise naturally in both the geometric and
analytic arguments. An existence theorem for weakly harmonic
1-fields, changing type on the unit circle, is derived under
Dirichlet conditions imposed on the non-characteristic portion of
the boundary. A similar system arises in the analysis of wave motion
near a caustic. A class of elliptic-hyperbolic boundary-value
problems is formulated for those equations as well. For both classes
of boundary-value problems, an arbitrarily small lower-order
perturbation of the equations is shown to yield solutions which are
strong in the sense of Friedrichs. MSC2000: 35M10, 58J32, 53A20,
78A05
\end{abstract}

\section{Introduction}

The projective disc was introduced by Beltrami$^3$ in 1868. His
construction was an early example of a Euclidean model for a
non-Euclidean space, in this case, a space having curvature equal
to $-1.$ The projective disc has the striking property that even
points infinitely distant from the origin are enclosed by the
Euclidean unit circle centered at the origin of $\mathbb{R}^{2}.$
This implies the possibility of points in projective space which
lie beyond the curve at infinity. It is known that such
\textit{ideal points} arise naturally in the process of
constructing normal and translated lines for chords of the
projective disc. In this sense ideal points may be said to be
intrinsic to the model, rather than only a theoretical possibility
allowed by the model. We call the union of the conventional
projective disc $\mathbb{P}^2$ and its ideal points the
\textit{extended projective disc.}

Hua$^9$ considered a second-order partial differential equation
for scalar functions on the extended projective disc. He proved
the existence of solutions to certain boundary-value problems of
\emph{Tricomi type,} in which data are given on characteristic
curves, which represent trajectories of generalized wavefronts.
Hua's work was extended to other problems of Tricomi type by Ji
and Chen.$^{10,11}$ The existence of a class of weak solutions to
the Hodge equations for harmonic 1-fields on extended
$\mathbb{P}^2,$ with data prescribed only on the
non-characteristic part of the boundary, was proven in Ref.\ 23.
Locally, the Hodge equations reduce in the smooth scalar case to
the equation studied by Hua.

This communication provides a geometric and analytic context for
such results (Sec.\ 1). In addition, we prove an existence theorem
for weakly harmonic 1-fields which includes the results of Ref.\
23 as a special case (Sec.\ 2.1), and consider a similar system
that arises in optics (Secs.\ 3.1, 3.2). Boundary-value problems
are formulated for both systems, in which the boundary contains
points in both the elliptic and hyperbolic regions of the
equations. These problems are shown in Secs.\ 2.2 and 3.3 to be an
arbitrarily small, lower-order perturbation away from problems
possessing a unique, strong solution.

Because both scalar equations and systems are discussed, we distinguish a
vector-valued solution by writing it in boldface. However, for typographic
simplicity, coefficient matrices and operators are not written in boldface.

\subsection{A geometric classification of linear second-order
operators}

The highest-order terms of any linear second-order partial
differential equation on a domain $\Omega \subset \mathbb{R}^{2}$
can be written in the form
\[
Lu=\alpha \left( x,y\right) u_{xx}+2\beta (x,y)u_{xy}+\gamma (x,y)u_{yy},
\]
where $(x,y)$ are coordinates on $\Omega $ and $\alpha ,$ $\beta ,$ and $%
\gamma $ are given functions. (A subscripted variable denotes partial
differentiation in the direction of the variable.)

If the discriminant
\[
\Delta \left( x,y\right) =\alpha \gamma -\beta ^{2}
\]
is positive, then the equation associated with the operator $L$ is said to
be of \textit{elliptic type}. The simplest example is Laplace's equation,
for which $\alpha =\gamma =1$ and $\beta =0.$ If the discriminant is
negative, \ then the equation associated with the operator $L$ is said to be
of \textit{hyperbolic type}. The simplest example is the normalized wave
equation, for which $\alpha =1,$ $\gamma =-1,$ $\beta =0;$ other forms are $%
\alpha =-1,$ $\gamma =1,$ $\beta =0,$ or $\alpha =\gamma =0,$ $\beta =1.$ If
$\Delta =0,$ then the equation associated with the operator $L$ is said to
be of \textit{parabolic type}; examples are equations which model diffusion.
If the discriminant is positive on part of $\Omega $ and negative elsewhere
on $\Omega ,$ then the equation associated with the operator $L$ is said to
be of \textit{mixed elliptic-hyperbolic type}. A simple example of an
elliptic-hyperbolic equation is the Lavrent'ev-Bitsadze equation, for which $%
\alpha =sgn\left( y\right) ,$ $\beta =0,$ and $\gamma =1.$

If we take $\Omega $ to be a smooth but curved surface, then we
may not be able to cover $\Omega $ by a single system of Cartesian
coordinates. However, we can always introduce Cartesian
coordinates $(x^{1},x^{2})$ \textit{locally} on any smooth
surface, in the neighborhood of a point on
the surface. In terms of such coordinates, the distance element $ds$ on $%
\Omega $ can be written in the form
\[
ds^{2}=\sum_{i=1}^{2}\sum_{j=1}^{2}g_{ij}(x^{1},x^{2})dx^{i}dx^{j},
\]
where $g_{ij}$ is a symmetric $2\times 2$ matrix, the \textit{metric tensor}
on $\Omega .$ (In the sequel we will understand repeated indices to have
been summed from 1 to $\dim \left( \Omega \right) $ without writing out the
summation notation each time.) A natural differential operator on functions $%
u$ defined on such a space is the \textit{Laplace-Beltrami operator}
\[
Lu=\frac{1}{\sqrt{\left| g\right| }}\frac{\partial }{\partial x^{i}}\left(
g^{ij}\sqrt{\left| g\right| }\frac{\partial u}{\partial x^{j}}\right) ,
\]
where $g^{ij}$ is the inverse of the matrix $g_{ij}$ and $g$ is its
determinant.

Laplace's equation can be associated to the Laplace-Beltrami operator on the
Euclidean metric for which $g_{ij}$ is the identity matrix. The wave
equation for $\beta =0$ can be associated with the Laplace-Beltrami operator
on the 2-dimensional Minkowski metric $g_{11}=1,$ $g_{22}=-1,$ $%
g_{12}=g_{21}=0.$ The Lavrent'ev-Bitsadze equation can be associated to the
Laplace-Beltrami operator on a metric which is Euclidean above the $x$-axis
and Minkowskian below the $x$-axis.

In this classification, the type of a linear second-order equation
is not a function of the associated linear operator at all; that
operator is always the Laplace-Beltrami operator. Rather, the type
of the equation is a feature of the metric tensor on an underlying
surface. A Riemannian metric, in which the distance between
distinct points of $\Omega $ is always positive, corresponds to an
elliptic equation, whereas a pseudoriemannian metric, for which
the distance between distinct points may be zero, corresponds to a
hyperbolic equation. The Laplace-Beltrami operator on a surface
for which the metric is Riemannian on part of a surface and
pseudoriemannian elsewhere will be of mixed elliptic-hyperbolic
type. However, any \textit{sonic} $-$ or \textit{parabolic} $-$
curve on which the change of type occurs will necessarily
represent a singularity of the metric tensor, as the determinant
$g$ will vanish along that curve. (The term \emph{sonic curve} is
borrowed from compressible fluid dynamics, in which the equations
for the velocity field of a steady ideal flow change from elliptic
to hyperbolic type at the speed of sound. The underlying
pseudoriemannian metric in that case is called the \emph{flow
metric}.$^4$)

One definition of the \textit{signature} of a metric is the sign of the
diagonal entries of the metric tensor. Any change in the signature which
results in a change in sign of the determinant $g$ will change the
Laplace-Beltrami operator on the metric from elliptic to hyperbolic type.
The Laplace-Beltrami operator on surface metrics for which such a change
occurs along a smooth curve will correspond to planar elliptic-hyperbolic
operators in local coordinates.

If we consider the distance element
\[
ds_{L}^{2}\equiv \alpha\left( x,y\right) dy^{2}-2\beta \left(
x,y\right) dxdy+\gamma \left( x,y\right) dx^{2},
\]
then \textit{null geodesics} on the corresponding surface are solutions of
the ordinary differential equation
\[
ds_{L}^{2}=0.
\]
The graphs of these solutions are called \textit{characteristic curves} of
the equation $Lu=0.$ Hyperbolic operators, which are associated with wave
propagation, always have real-valued \emph{characteristics,} or null
geodesics.

In determining the qualitative behavior of solutions to partial
differential equations we often ignore lower-order terms, but this
neglect is only justified when considering purely second-order
properties such as the nature of the sonic curve. The importance
to this paper of lower-order terms is related to the fact that the
Laplace-Beltrami equations on the extended projective disc are not
of \textit{real principal type} in the sense of Ref.\ 6; see Ref.\
27 for an accessible discussion of scalar elliptic-hyperbolic
operators of real principal type and their properties.

\subsection{The geometry and analysis of ideal points}

Here we review basic properties of Laplace-Beltrami equations on
Beltrami's hyperbolic metric on the projective disc:
\[
ds^{2}=\frac{\left( 1-y^{2}\right) dx^{2}+2xydxdy+\left(
1-x^{2}\right) dy^{2}}{\left( 1-x^{2}-y^{2}\right) ^{2}}
\]
(see , \emph{e.g.,} Ref.\ 32, Vol. I, Sec.\ 65 and Vol. II, Sec.\
138, for a derivation). In this metric the unit circle is the
\emph{absolute}: the locus of points at infinity.

The existence of points lying beyond the curve at infinity on the
projective disc is natural from a geometric point of view. For
example, choose a point $p$ in the interior of the projective disc
and draw a vertical line $\ell_v$ through it. A \textit{hyperbolic
line} in the Beltrami metric is any open chord of the unit circle,
so $\ell_v$ is a hyperbolic line plus two points at infinity and
an \textit{ideal extension} to points outside the unit circle.
Denote by $F(p)$ the family of hyperbolic lines created by
rotating $\ell_v$ about $p.$ Move $p$ along the horizontal line
$\ell _{h}$ through $p,$ and consider the affect of this motion on
the family $F(p).$ As $p$ passes through the boundary of the unit
circle $\kappa$ into the $\mathbb{R}^{2}$-complement of $\kappa,$
the family of hyperbolic rotations becomes a family of hyperbolic
translations. For this reason, hyperbolic translations inside the
unit disc can be interpreted as rotations about a point in
$\mathbb{R}^{2}$ lying beyond the unit disc.

As another example, consider that the \textit{pole} of a hyperbolic
line $\ell $ is the intersection of those two tangents to the unit
circle which intersect $\ell$ at the two points of its contact with
the unit circle. (We call these the \emph{polar lines} of $\ell$.)
Thus any two hyperbolic lines $\ell _{1}$ and $\ell _{2}$ are
orthogonal if and only if the pole of $\ell _{2}$ lies on the ideal
extension of $\ell _{1}$ and \textit{vice-versa}.

These and other geometric constructions on extended
$\mathbb{P}^{2}$ are described in more detail in Chapter 4 of
Ref.\ 28.

In order to see that ideal points also arise naturally in analysis,
consider the Laplace-Beltrami operator on the projective disc with
Beltrami's metric. We have
\[
L\left[ u\right] =\left( 1-x^{2}-y^{2}\right) [ \left( 1-x^{2}\right)
u_{xx}-2xyu_{xy}+\left( 1-y^{2}\right) u_{yy}
\]

\[
+lower\;order\;terms].
\]

The characteristics of the equation $L[u]=0$ satisfy the ordinary
differential equation
\begin{equation}
\left( 1-y^{2}\right) dx^{2}+2xydxdy+(1-x^{2})dy^{2}=0.
\end{equation}
This equation has solutions
\begin{equation}
x\cos \theta +y\sin \theta =1,
\end{equation}
where, as is conventional, we take $\theta $ to be the angle
between the radial vector and the positive $x$-axis. Solutions of
eq.\ (2) correspond geometrically to the family of tangent lines
to the unit circle centered at the origin of $\mathbb{R}^{2}.$

Thus the characteristic lines always include ideal points and wave
propagation can only occur on regions composed of such points.

The Laplace-Beltrami equations on extended $\mathbb{P}^{2}$ come
with a natural gauge theory in the following sense: The
characteristic equation is obviously invariant under the
projective group. So although the equations in
the form in which we study them change type on the unit circle in $\mathbb{R}%
^{2},$ they are projectively equivalent to a system which changes
type on any conic section. Note that whereas classical gauge
theories are invariant under groups of Euclidean motions, which
are inertial transformations, this kind of gauge invariance is
with respect to a group of non-Euclidean motions, which are
non-inertial. Also, the gauge theories which are familiar from
particle physics act ``upstairs'' on a fiber bundle of physical
states. The transformation group under which the Laplace-Beltrami
equations are invariant acts ``downstairs'' on the underlying
metric, in the manner of the gauge group of general relativity.
Indeed, analysis of wave motion on extended $\mathbb{P}^2$ has
certain similarities to the analysis of wave motion in the
vicinity of a light cone (\emph{c.f.} Ref.\ 30). The time-like and
space-like regions are inverted, and characteristic lines for the
Laplace-Beltrami equation are analogous to the paths of photons.

\section{Harmonic 1-fields on the extended projective disc}

We can solve, instead of the Laplace-Beltrami equation, a system of two
first-order equations of the form
\begin{equation}
\left| g\right| ^{-1/2}\partial _{i}\left( g^{ij}\sqrt{\left| g\right| }%
u_{j}\right) =0,
\end{equation}
\begin{equation}
\frac{1}{2}\left( \partial _{i}u_{j}-\partial _{j}u_{i}\right) =0,
\end{equation}
where $u_i=u_i(x^1,x^2),$ $i=1,2.$ As in the second-order equation, $g_{ij}$
is a metric tensor on the underlying surface. Solutions $\mathbf{u}%
=(u_{1},u_{2})$ of this first-order system are (locally) \textit{harmonic
1-fields.} Notice that if the scalar function $\varphi \left( x^1,x^2\right)
$ satisfies $\varphi _{x^1}=u_{1}$ and $\varphi _{x^2}=u_{2},$ then $\varphi
$ satisfies the Laplace-Beltrami equations. But there are solutions $\varphi
$ of the Laplace-Beltrami system for which the pair $\left( \varphi
_{x^1},\varphi _{x^2}\right) $ is not a harmonic 1-field.

Consider a system of first-order equations on $\mathbb{R}^{2}$
having the form
\begin{equation}
L\mathbf{u}=\mathbf{f},
\end{equation}
where
\[
L=\left( L_{1},L_{2}\right) ,\;\mathbf{f}=\left( f_{1},f_{2}\right) ,
\]
\[
\mathbf{u}=\left( u_{1}\left( x,y\right) ,u_{2}\left( x,y\right)
\right) ,\;\left( x,y\right) \in \Omega \subset \subset
\mathbb{R}^{2}.
\]
Let $\mathbf{u}$ satisfy (5) with
\begin{equation}
\left( L\mathbf{u}\right) _{1}=\left[ \left( 1-x^{2}\right)
u_{1}\right] _{x}-2xyu_{1y}+\left[ \left( 1-y^{2}\right)
u_{2}\right] _{y}+k_1xu_{1}+k_2yu_{2},
\end{equation}
and
\begin{equation}
\left( L\mathbf{u}\right) _{2}=\left( 1-y^{2}\right) \left(
u_{1y}-u_{2x}\right) +k_3xu_1+k_4yu_2 ,
\end{equation}
where $\Omega$ is chosen so that $y^2 \ne 1$ there. Here
$k_1,k_2,k_3$ and $k_4$ are constants representing lower-order
coefficients. In this section we consider three particular
distributions of lower-order terms, studied in Ref.\ 23:

\emph{Case 1}: $k_1=k_2=-2,\,k_3=k_4=0.$ The domain of eqs.\
(5)-(7) in this case will be called $\Omega_1.$

\emph{Case 2}: $k_1=-2,\,k_2=k_3=0,\,k_4=2.$ The domain of eqs.\
(5)-(7) in this case will be called $\Omega_2.$

\emph{Case 3}: $k_1=k_2=k_3=k_4=0.$ The domain of eqs.\ (5)-(7) in
this case will be called $\Omega_3.$ This case corresponds to
eqs.\ (3), (4).

The union of the domains $\Omega_1,$ $\Omega_2,$ and $\Omega_3$
will be called $\Omega.$

A system of first-order equations can also be said to be of
elliptic or hyperbolic type, and thus may change type along a
singular curve. See, \textit{e.g.,} Ref.\ 5, Ch.\ III.2. The
higher-order terms of the preceding system can be written in the
form $A^1\mathbf{u}_{x}+A^2\mathbf{u}_{y}, $ where
\begin{equation}
A^1=\left[
\begin{array}{cc}
1-x^{2} & 0 \\
0 & -\left( 1-y^{2}\right)
\end{array}
\right]
\end{equation}
and
\begin{equation}
A^2=\left[
\begin{array}{cc}
-2xy & 1-y^{2} \\
1-y^{2} & 0
\end{array}
\right] .
\end{equation}

If $y^{2}\ne 1,$ the characteristic equation
\[
\left| A^1-\lambda A^2\right| =-\left( 1-y^{2}\right) \left[
\left( 1-y^{2}\right) \lambda ^{2}+2xy\lambda +\left(
1-x^{2}\right) \right]
\]
possesses two real roots $\lambda _{1},\lambda _{2}$ on $\Omega $ precisely
when $x^{2}+y^{2}>1$. \ Thus the system is elliptic in the intersection of $%
\Omega $ with the open unit disc centered at $\left( 0,0\right) $ and
hyperbolic in the intersection of $\Omega $ with the complement of the
closure of this disc. The boundary of the unit disc, along which this change
in type occurs, is the line at infinity on the projective disc and a line
singularity of the tensor $g_{ij}.$

Denote by $\Omega $ a region of the plane for which part of the boundary $%
\partial \Omega $ consists of a family of curves $\Gamma $ composed of
points satisfying eq.\ (1) and the remainder $C=\partial \Omega
\backslash \Gamma $ of the boundary consists of points $(x,y)$
which do not satisfy eq.\ (1). We seek solutions of eqs.\ (5)-(7)
which satisfy the boundary condition
\begin{equation}
u_{1}\frac{dx}{ds}+u_{2}\frac{dy}{ds}=0,
\end{equation}
where $s$ denotes arc length, on the non-characteristic part $C$ of the
domain boundary. Because the tangent vector $\mathbf{T}$ on $C$ is given by
\[
\mathbf{T}=\frac{dx}{ds}\mathbf{i}+\frac{dy}{dx}\mathbf{j},
\]
a geometric interpretation of this boundary condition is that the dot
product of the vector $\mathbf{u}=(u_{1},u_{2})$ and the tangent vector to $%
C $ vanishes, \emph{i.e.}, $\mathbf{u}$ is normal to the boundary $\partial
\Omega $ on the boundary section $C$. We call these \emph{homogeneous
Dirichlet conditions}.

\subsection{Weak solutions}

In Ref.\ 23, weak solutions to (5)-(7), (10) are shown to exist in
certain weighted $L^{2}$ spaces on a class of domains. Here we
extend that result to the case in which the domain is formed by
the polar lines of a hyperbolic line $\ell $ and a smooth curve
$C$ extending between the two polar lines of $\ell.$ The curve $C$
must have the property that $dy_{|C}\leq 0$ when $\partial \Omega
$ is traversed in a counterclockwise direction. However, as long
as this condition is met, $C$ need not intersect the polar lines
of $\ell$ at their points of tangency with the unit circle. Thus
$C$ may extend into both the elliptic and the hyperbolic regions.

This domain is the analogue of the ``ice-cream cone''-shaped
domain associated to the \emph{Tricomi equation}$^{31}$
\[
yu_{xx}+u_{yy}=0,
\]
where in our case the curve $C$ is the boundary of the ice-cream
part and the polar lines, which are characteristics of eqs.\
(5)-(7), are the boundary of the cone part. The unit circle is the
analogue of the $x$-axis, which is the sonic curve for the Tricomi
equation.

We initially consider the distribution of lower-order terms in
case 1 of eqs.\ (6), (7). Let $\theta$ lie in the interval
$\lbrack 0,\pi /4 \rbrack$ and denote by $\Omega_1$ the region of
the first and fourth quadrants bounded by the characteristic line
\[
\Gamma _{1}:x\cos \theta + y\sin \theta =1,
\]
the characteristic line
\[
\Gamma _{2}:x\cos \theta - y\sin \theta =1,
\]
and a smooth curve $C.$ Let $C$ intersect the lines $\Gamma_1,$ $%
\Gamma_2$ at two distinct points $c_1,$ $c_2,$ respectively.
Assume that $\forall \left( x,y\right) \in \Omega_1,$
$1/\sqrt{2}\leq x<\sqrt{2}$ and $-1/\sqrt{2} \leq y<1/\sqrt{2},$
and that $dy\leq 0$ on $C.$ A cusp is permitted for $\theta =
\pi/4$ at the points $c_1,c_2=(1/\sqrt{2}, \pm 1/\sqrt{2}).$
Otherwise, the boundary will have piecewise continuous tangent (so
that Green's Theorem can be applied to it). Note that the domain
considered in Sec.\ 3 of Ref.\ 23 is equivalent to this domain in
the degenerate special case $\theta =0.$

Define $U$ to be the vector space consisting of all pairs of measurable
functions $\mathbf{u}=\left( u_{1},u_{2}\right) $ for which the weighted $%
L^{2}$ norm
\[
\left\| \mathbf{u}\right\| _{\ast }=\left[ \int \int_{\Omega_1
}\left( \left| 2x^{2}-1\right| u_{1}^{2}+\left| 2y^{2}-1\right|
u_{2}^{2}\right) dxdy\right] ^{1/2}
\]
is finite. Notice that this expression vanishes at the
intersection of $\ell$ with its polar lines at the value
$\theta=\pi/4.$ Denote by $W$ the linear space defined by pairs of
functions $\mathbf{w}=\left( w_{1},w_{2}\right) $ having
continuous derivatives and satisfying:
\begin{equation}
    w_{1}dx+w_{2}dy=0
\end{equation}
on $\Gamma = \Gamma_1 \bigcup \Gamma_2$;
\begin{equation}
    w_{1}=0
\end{equation}
on $C;$ and
\[
\int \int_{\Omega_1 }\left[ \left| 2x^{2}-1\right| ^{-1}\left( L^{\ast }%
\mathbf{w}\right) _{1}^{2}+\left| 2y^{2}-1\right| ^{-1}\left( L^{\ast }%
\mathbf{w}\right) _{2}^{2}\right] dxdy<\infty .
\]
Here
\[
\left( L^{\ast }\mathbf{w}\right) _{1}=\left[ \left( 1-x^{2}\right) w_{1}%
\right] _{x}-2xyw_{1y}+\left[ \left( 1-y^{2}\right) w_{2}\right]
_{y}+2xw_{1},
\]
and
\[
\left( L^{\ast }\mathbf{w}\right) _{2}=\left( 1-y^{2}\right) \left(
w_{1y}-w_{2x}\right) +2yw_{1}.
\]
Define the Hilbert space $H$ to consist of pairs of measurable functions $%
\mathbf{h}=\left( h_{1},h_{2}\right) $ for which the norm
\[
\left\| \mathbf{h}\right\| ^{\ast }=\left[ \int \int_{\Omega_1
}\left( \left| 2x^{2}-1\right| ^{-1}h_{1}^{2}+\left|
2y^{2}-1\right| ^{-1}h_{2}^{2}\right) dxdy\right] ^{1/2}
\]
is finite.

We say that $\mathbf{u}$ is a \textit{weak solution} of the system
(5)-(7), (10) in case 1 on $\Omega_1 $ if $\mathbf{u}\in U$ and
for every $\mathbf{w}\in W,$

\[
-\left( \mathbf{w},\mathbf{f}\right) =\left( L^{\ast }\mathbf{w},\mathbf{u}%
\right) ,
\]
where
\[
\left( \mathbf{w}, \mathbf{f}\right) =\int \int_{\Omega_1 }\left(
w_{1}f_{1}+w_{2}f_{2}\right) dxdy.
\]

In case 2, we restrict the domain $\Omega_2$ to lie in the fourth
quadrant of the Cartesian plane. Define $\Gamma_1, \,\Gamma_2
\subset \Gamma$ to be characteristic lines which are tangent to
the unit circle at distinct points in the fourth quadrant and
which intersect at a point in the complement of the unit disc in
$\mathbb{R}^2.$ The curve $C$ is defined analogously to the
corresponding curve of $\Omega_1.$ In particular, $dy_{|C} \leq 0$
on $\Omega_2$ when $C$ is traversed in a counter-clockwise
direction. Replace $U$ by the space $U'$ of all pairs $\textbf{u}$
of measurable functions $\left(u_1,u_2\right)$ for which the
weighted $L^2$ norm

\[
\left\| \mathbf{u}\right\|' _{\ast }=\left[ \int
\int_{\Omega_2}\left(  x  u_{1}^{2}+\left| y\right|
u_{2}^{2}\right) dxdy\right] ^{1/2}
\]
is finite. Replace $W$ by the space $W'$ defined by pairs of
continuously differentiable functions $\mathbf{w}=\left(
w_{1},w_{2}\right) $ satisfying eq.\ (11) on $\Gamma,$ eq.\ (12)
on $C,$ and
\[
\int \int_{\Omega_2 }\left[  x^{-1}\left( L^{\ast }%
\mathbf{w}\right) _{1}^{2}+\left| y\right| ^{-1}\left( L^{\ast }%
\mathbf{w}\right) _{2}^{2}\right] dxdy<\infty .
\]
In this case
\[
\left( L^{\ast }\mathbf{w}\right) _{1}=\left[ \left( 1-x^{2}\right) w_{1}%
\right] _{x}-2xyw_{1y}+\left[ \left( 1-y^{2}\right) w_{2}\right]
_{y}+2xw_{1},
\]
and
\[
\left( L^{\ast }\mathbf{w}\right) _{2}=\left( 1-y^{2}\right)
\left( w_{1y}-w_{2x}\right) -2yw_{2}.
\]
Finally, we replace $H$ by the space $H'$ of measurable functions $%
\mathbf{h}=\left( h_{1},h_{2}\right) $ for which the norm
\[
\left\| \mathbf{h}\right\|{'}^{\ast }=\left[
\int\int_{\Omega_2}\left( x^{-1}h_{1}^{2}+\left| y \right|
^{-1}h_{2}^{2}\right) dxdy\right] ^{1/2}
\]
is finite.

Because $k_4$ is nonzero in case 2, the consistency condition (4)
is violated and $\mathbf{u}$ cannot be the gradient of a scalar
potential, even locally. Harmonic fields in which condition (4) is
violated arise in various contexts $-$ see Section 4 of Ref.\ 25
for a nonlinear example $-$ and correspond physically to
stationary fields having sources.

In case 3, we restrict the domain, $\Omega_3,$ to lie in the first
quadrant. Define $\Gamma_1,\, \Gamma_2 \subset \Gamma$ to be
characteristic lines which are tangent to the unit circle at
distinct points in the first quadrant and which intersect at a
point in the complement of the unit disc in $\mathbb{R}^2.$ In
this case we replace $U$ and $H$ by $L^2.$ We replace $W$ by the
space of pairs of $L^2$ functions $(w_1,w_2)$ which satisfy (11)
on $\Gamma$ and (12) on $C.$ Note that $L$ is self-adjoint in case
3. In addition, we fix positive numbers $\delta << 1/2$ and
$\varepsilon << 1/2$ and require $\Omega_3$ to lie in the
semi-infinite rectangle

\[
\frac{1}{\sqrt{2}}<x,\,\frac{1}{\sqrt{2-\delta}}<y\leq
\sqrt{1-\varepsilon}.
\]

Weak solutions in cases 2 and 3 are defined exactly analogously to
case 1, with appropriate replacement of the domain and function
spaces.

\bigskip

\textbf{Theorem 1}. \emph{Let the lower-order terms in eqs.\ (6),
(7) be distributed as in cases 1, 2, or 3, on the domains
$\Omega_1,$ $\Omega_2,$ or $\Omega_3,$ respectively. Then there
exists a weak solution of the boundary-value problem (5)-(7), (10)
for every $\mathbf{f}\in H.$}

\bigskip

\textit{Proof}. The proof is an extension of the arguments in
Ref.\ 23, so we will be brief. We derive a \textit{basic
inequality}, that there is a $\,K\in \mathbb{R}^{+}$ such that
$\forall \mathbf{w}\in W,$

\[
K\left\| \mathbf{w}\right\| _{\ast }\leq \left\| L^{\ast
}\mathbf{w}\right\| ^{\ast }
\]
(with the norms appropriately adjusted in cases 2 and 3). We
derive this inequality by choosing a scalar multiplier $a,$
computing the $L^{2}$ inner product $\left( L^{\ast }\mathbf{w},a
\mathbf{w}\right),$ and integrating by parts. Denoting the
coefficients of $w_{1}^{2}$ off the boundary by $\alpha ,$ those
of $w_{2}^{2}$ by $\gamma $ and those of $w_{1}w_{2}$ by $2\beta,
$ we choose, in case 1, $a=x^{2}$ and obtain
\[
\alpha =x\left( 3x^{2}-1\right),\, \gamma =x\left( 1-y^{2}\right),
\]
and
\[
\beta =yx^{2},
\]
where
\[
2\beta w_{1}w_{2}\geq -2x\left| xw_{1}\right| \left| yw_{2}\right| \geq
-\left( x^{3}w_{1}^{2}+xy^{2}w_{2}^{2}\right) .
\]

In case 2 we choose $a=1$ and obtain
\[
\alpha = 2x, \, \gamma=-2y,
\]
and $\beta =0.$

In case 3 we choose $a=xy$ and obtain
\[
\alpha = \frac{y}{2}\left(3x^2-1\right), \, \gamma =
\frac{y}{2}\left(1-y^2\right),
\]
and
\[
2\beta = -\left(1-y^2\right)x.
\]
The quadratic form $\alpha\gamma-\beta^2$ can be shown to be
non-negative in case 3 by noticing that the argument in Sec.\ 6.2
of Ref.\ 23 does not use the restriction $x \leq 1$ and thus
extends to our more general case.

The remainder of the proof is essentially the same for all three
cases. Applying Green's Theorem to derivatives of products in
$\left( L^{\ast }\mathbf{w},a \mathbf{w}\right),$ we obtain a
boundary integral $I$ having the form

\[
\int_{\partial \Omega} \frac{a} {2} \left[ (1-x^2)w_1^2 dy+2xy w_1^2 dx %
\right]
\]
\[
-\int_{\partial \Omega} a \left[ (1-y^2)w_1w_2dx+
\frac{1}{2}(1-y^2) w_2^2 dy \right].
\]
Because $w_1$ vanishes identically on $C,$ the boundary integral
is nonnegative on $C$ by the hypothesis on $dy_{|C}.$ On the
characteristic curves, we no longer have the property that $dx=0,$
which we used in deriving the basic inequality of Ref.\ 23.
However,

\[
I_{|\Gamma} = \int_\Gamma \frac {a} {2}
\left\{(1-x^2)w_1^2dy+[2xyw_1^2-(1-y^2)w_1w_2]dx \right\},
\]
where we have used the fact that

\[
w_2dy=-w_1dx
\]
on characteristic lines. In fact, we have

\[
I_{|\Gamma} = \int_{\Gamma} \frac{a}{2}\left[(1-x^2)w_1^2\left(\frac{dy}{dx%
}\right)+2xyw_1^2-(1-y^2)w_1w_2\right]dx
\]
\[
= \int_{\Gamma} \frac{a}{2}\left[-(1-x^2)w_1w_2\left(\frac{dy}{dx}%
\right)^2+2xyw_1^2-(1-y^2)w_1w_2\right]dx
\]
by the same identity. Equation (1) implies that

\[
-(1-x^2)\left(\frac{dy}{dx}\right)^2=2xy\frac{dy}{dx}+1-y^2,
\]
so we can write

\[
I=\int_{\Gamma }\frac{a}{2}\left[2xy\frac{dy}{dx} + 1 -
y^{2}\right] w_{1}w_{2}dx+
\]
\[
\int_{\Gamma }\frac{a}{2} \left[
2xyw_{1}(-w_{2}\frac{dy}{dx})-(1-y^{2})w_{1}w_{2}\right] dx=0.
\]
This establishes the basic inequality.

Proceeding as in Ref.\ 19, we use the basic inequality to apply
the Riesz Representation Theorem and obtain an element
$\mathbf{h}\in H $ for which

\[
-(\mathbf{w},\mathbf{f})=-(L^{\ast }\mathbf{w},\mathbf{h})^{\ast},
\]
where the product on the right is the inner product on $H$ (or on
$H'$ or $L^2$ in cases 2 or 3, respectively). Writing $h_1$ and
$h_2$ of $\textbf{h}$ in terms of appropriate rescalings of $u_1$
and $u_2,^{23}$ we obtain
\[
-(L^{\ast}\mathbf{w},\mathbf{h})^{\ast}=(L^{\ast}\mathbf{w}, \mathbf{u}),
\]
which completes the proof.

\subsection{Strong solutions}

By a \textit{strong solution} of the boundary-value problem (5),
(10) we mean an element $\textbf{u}\in L^2(\Omega)$ for which
there exists a sequence $\textbf{u}^{\nu }$ of continuously
differentiable vectors satisfying the boundary condition (10), for
which
\[
\lim_{\nu \rightarrow \infty }\left\| \textbf{u}^{\nu
}-\textbf{u}\right\|_{L^2} =0,
\]
and
\[
\lim_{\nu \rightarrow \infty }\left\| L\textbf{u}^{\nu
}-\textbf{f}\right\|_{L^2} =0.
\]

For $\mathbf{u}=(u_{1}(x,y),u_{2}(x,y))$, $(x,y)\in \Omega \subset
\subset \mathbb{R}^{2}$, define the operator $L=(L_{1},L_{2})$ by
the matrix equation

\begin{equation}
L\mathbf{u}=A^1\mathbf{u}_x+A^2\mathbf{u}_y+B\mathbf{u}
\end{equation}
for matrices $A^1$, $A^2$, and $B$. We say that $L$ is
\emph{symmetric-positive}$^{7,15,16}$ if the matrices $A^1$ and
$A^2$ are symmetric and the matrix

\[
    Q \equiv 2B^* - A_x^1 - A_y^2
\]
is nonnegative. Here

\[
B^* = \frac{1}{2} (B+B^t),
\]
where for a matrix $W=[w_{ij}]$, $W^t=[w_{ji}]$.

In cases for which $L$ is not symmetric-positive, there may be a
nonsingular matrix $E$ such that $EL$ is symmetric-positive. In
that case we replace the equation
\[
L\textbf{u}=\textbf{f}
\]
by the equation
\[
EL\textbf{u}=E\textbf{f}
\]
and try to show that the operator $EL$ is symmetric-positive. (The
conversion of $L$ into a symmetric-positive operator by the
construction of a suitable multiplier $E$ will not be used in this
section, but will be used in Sec.\ 3.3.)

Suppose that $N(x,y),\,(x,y)\in\partial\Omega,$ is a linear
subspace of the vector space $V,$ where $\textbf{u}$ is regarded
as a mapping $\textbf{u}:\Omega \cup \partial \Omega \rightarrow
V,$ and that $N(x,y)$ depends smoothly on $x$ and $y.$ Define the
matrix
\[
\beta = n_1A_{|\partial \Omega}^1 + n_2A_{|\partial \Omega}^2,
\]
where $\textbf{n}=\left(n_1,n_2\right)$ is the outward-pointing
normal vector to $\partial \Omega.$ The boundary condition that
$u$ lie in $N$ is said to be \emph{admissible}$^{15}$ if $N$ is a
maximal subspace of $V$ and if the quadratic form
$(\textbf{u},\beta \textbf{u})$ is non-negative on $\partial
\Omega.$

A sufficient condition$^7$ for admissibility is that there exist a
decomposition
\[
\beta = \beta_++\beta_-,
\]
for which the direct sum of the null spaces for $\beta_+$ and
$\beta_-$ spans the restriction of $V$ to the boundary, the
intersection of the ranges of $\beta_+$ and $\beta_-$ have only the
vector $\textbf{u}=0$ in common, and the matrix
$\mu=\beta_+-\beta_-$ satisfies
\[
\mu^\ast = \frac{\mu + \mu^t}{2} \geq 0.
\]
In this case the boundary condition
\[
\beta_-\textbf{u}=0 \, on \, \partial \Omega
\]
is admissible for the boundary-value problem
\[
L\textbf{u}=f \, in \, \Omega.
\]
Moreover, the boundary condition
\[
\beta_+^t \textbf{w}=0 \, on \, \partial \Omega
\]
is admissible for the adjoint problem
\[
L^\ast \textbf{w}=\textbf{h} \, in \, \Omega.
\]
These two problems possess unique, strong solutions whenever the
differential operators are symmetric-positive and the boundary
conditions are admissible.$^{7,15}$

In this section we give sufficient conditions for the existence of
certain strong solutions arising from an arbitrarily small
lower-order perturbation of the Laplace-Beltrami equations on
extended $\mathbb{P}^2.$ We do so by showing that the differential
operator $L$ given by (5)-(7) with $k_1=k_2=k_3=k_4=0$ is
arbitrarily close to a symmetric-positive operator and by stating
an admissible boundary condition. The existence of strong
solutions to a different perturbation on an explicit domain will
be shown in Sec.\ 3.3.

If the matrices $A^1$ and $A^2$ of eq.\ (13) are given by eqs.\
(8) and (9) and the matrix $B$ is given by
\[
\left(%
\begin{array}{cc}
  -2x & -2y \\
  0 & 0 \\
\end{array}%
\right),
\]
then the quantity $Q$ is zero. Thus we replace the matrix $B$ by a
matrix $B_\varepsilon$ which differs from $B$ by an arbitrarily
small perturbation and takes the form

\begin{equation}
    B_\varepsilon = \left(%
\begin{array}{cc}
  -2x +\varepsilon_1 & -2y +\varepsilon_2\\
  \left(1-y^2\right)\varepsilon_3 & \left(1-y^2\right)\varepsilon_4 \\
\end{array}%
\right),
\end{equation}
where $\varepsilon_1>0,$ $\varepsilon_4 >0,$
$\varepsilon_2+\left(1-y^2\right)\varepsilon_3\geq 0,$ and

\[
\left[\varepsilon_2+\left(1-y^2\right)\varepsilon_3\right]^2\leq
4\left(1-y^2\right)\varepsilon_1\varepsilon_4.
\]
If we choose the domain of $L$ in such a way that $y^2 < 1$ there,
then this replacement converts $Q$ into a positive-definite matrix
and $L$ into a symmetric-positive operator.

Denote by $\Omega_4$ a domain having $C^2$ boundary $\partial
\Omega_4=E \cup F$ such that $y^2 < 1$ on $\Omega_4.$ Let the
components of the normal vector $\textbf{n}$ on $\partial
\Omega_4$ be given by $(n_1,n_2).$ Assume that $n_1$ and $n_2$
never vanish at the same point of $\partial \Omega_4.$ We place
conditions on $n_1,$ $n_2,$ and $\partial \Omega_4$ sufficient to
guarantee admissibility of the boundary condition

\begin{equation}
    u_1n_2-u_2n_1=0
\end{equation}
on $F$, with no condition given on $E.$

Let $n_1 \geq 0,$ $n_2 \leq 0$ on $F$ and $n_1 \leq 0,$ $n_2 \geq
0$ on $E.$ Defining the adjoint space as in Sec.\ 2.1, for
$\textbf{w} \in V^{\ast}$ we take $\textbf{w}=(0,w_2)$ on $F$ and

\[
w_1n_2-w_2n_1=0
\]
on $E.$ Define

\[
\alpha=\left[-\left(1-y^2\right)n_2/n_1+2xy-\left(1-x^2\right)n_1/n_2\right]n_2.
\]
Assume that $\alpha=0$ on $E,$ and that $\alpha \leq 0$ on $F.$

\bigskip

\textbf{Theorem 2}. \emph{The boundary-value problem}
\[
L\textbf{u} = A^1\textbf{u}_x+A^2\textbf{u}_y+B_\varepsilon
\textbf{u} = \textbf{f}
\]
\emph{for $(x,y) \in \Omega_4$, with $A^1$, $A^2,$ and
$B_\varepsilon$ given by eqs.\ (8), (9), and (14) respectively and
with condition (15) imposed on the curve $F$ of $\partial
\Omega_4,$ possesses a unique, strong solution} $\textbf{u}(x,y)$
\emph{for every} $\textbf{f} \in L^2(\Omega_4).$

\bigskip

\emph{Proof}. Because the matrix $B_\varepsilon$ has been
constructed in such a way that $L$ is symmetric-positive, it
remains only to show that the boundary condition (15) is
admissible on $\Omega_4.$

We have

\[
\beta = \left(%
\begin{array}{cc}
  -\alpha - \left(1-y^2\right)n_2^2/n_1 & \left(1-y^2\right)n_2 \\
  \left(1-y^2\right)n_2 & -\left(1-y^2\right)n_1 \\
\end{array}%
\right).
\]
Note that the apparent singularities in $\beta$ at $n_1=0$ and in
$\alpha$ at $n_2=0$ are removable.

On $F,$ choose

\[
\beta_+=\left(%
\begin{array}{cc}
  -\alpha & 0 \\
  0 & 0 \\
\end{array}%
\right)
\]
and

\[
\beta_-=\left(1-y^2\right)n_2\left(%
\begin{array}{cc}
  -n_2/n_1 & 1 \\
  1 & -n_1/n_2 \\
\end{array}%
\right).
\]
On $E,$ choose

\[
\beta_+=\left(1-y^2\right)n_2\left(%
\begin{array}{cc}
  -n_2/n_1 & 1 \\
  1 & -n_1/n_2 \\
\end{array}%
\right)
\]
and

\[
\beta_-=\left(%
\begin{array}{cc}
  -\alpha & 0 \\
  0 & 0 \\
\end{array}%
\right).
\]

If $\textbf{u} \in V_{|F},$ then (15) implies that
$\beta_-\textbf{u}=0.$ The properties of $V^{\ast}$ imply that
$\textbf{w}^t\beta_+=0$ for $\textbf{w} \in V^{\ast}_{|F}.$ If
$\textbf{u} \in V_{|E},$ then $\beta_-\textbf{u}=0$ for all values
of $\textbf{u}$ and $\textbf{w}^t\beta_+=0$ by the properties of
$V^{\ast}$ and $\alpha.$ So the direct sum of the null spaces of
$\beta_-$ and $\beta_+$ spans $V$ on $\partial \Omega_4.$
Moreover, the hypotheses guarantee that the ranges of $\beta_-$
and $\beta_+$ have only the zero vector in their intersection.
Finally,

\[
\beta_+-\beta_- = \mu^{\ast}\geq 0
\]
on both $E$ and $F.$

This completes the proof of Theorem 2.

\section{An analogous problem from optics}

Geometrical optics is a zero-wavelength approximation to classical
wave mechanics in which the governing differential equations are
replaced by the Euclidean geometry of rays. The limitations of the
geometrical optics approximation are apparent in the neighborhood
of \emph{caustics}, which are envelopes of a family of rays. It is
not simply that geometrical optics predicts infinite intensity in
such regions, whereas diffractive effects reduce the predicted
intensity to a finite number. Even in applications for which the
agreement between the predictions of geometrical optics and
experiment is generally good, the former may predict
singularities, \textit{e.g.,} cusps, which are entirely smoothed
out by diffraction. A dramatic example of this for the case of
water waves is illustrated in Figures 5.6.1 and 5.6.2 of Ref.\ 29.
This is, of course, far from the only drawback of the geometrical
optics approximation. See, for example, the discussion of the
rainbow caustic in Sec.\ 6.3 of Ref.\ 22.

The accuracy of the geometrical optics approximation can be
improved by considering waves of arbitrarily high frequency
obtained by uniform asymptotic approximation of solutions to the
Helmholtz equation (Sec.\ 3.1). While the older of these
approximations also fail at caustics, an asymptotic formula
introduced independently by Kravtsov$^{12}$ and Ludwig$^{17}$
retains its meaning even in the neighborhood of a caustic; see
Ref.\ 13 for a review.

Recently, Magnanini and Talenti studied a nonlinear
elliptic-hyperbolic equation, implied by the Ludwig-Kravtsov
approximation, having the form$^{18}$
\begin{equation}
\left( \left| \nabla v\right| ^{4}-v_{y}^{2}\right)
v_{xx}+2v_{x}v_{y}v_{xy}+\left( \left| \nabla v\right| ^{4}-v_{x}^{2}\right)
v_{yy}=0,
\end{equation}
where $v=v(x,y),$ $\left( x,y\right) \in \mathbb{R}^{2}.$ Those
authors were able to show the existence of weak solutions to the
full Dirichlet problem for the linear elliptic-hyperbolic equation
\begin{equation}
\left[ \left( p^{2}+q^{2}\right) ^{2}-p^{2}\right] V_{pp}-2pqV_{pq}+\left[
\left( p^{2}+q^{2}\right) ^{2}-q^{2}\right] V_{qq}=0,
\end{equation}
which is related to eq.\ (16) by the \textit{Legendre
transformation}
\begin{equation}
V_L(p,q)=xp+yq-v(x,y).
\end{equation}

Magnanini and Talenti's result is remarkable in that it is
difficult to formulate a full Dirichlet problem which is
well-posed for a given elliptic-hyperbolic equation, even in the
weak sense; by \textit{full} we mean that data are prescribed on
the entire boundary. Morawetz's proof of the existence of weak
solutions to the full Dirichlet problem for the Tricomi equation,
the most intensively studied elliptic-hyperbolic equation,
required a delicate argument.$^{20,27}$ The full Dirichlet
problems for other important elliptic-hyperbolic equations remain
unknown. For example, the full Dirichlet problem has not been
correctly formulated even for weak solutions to a scalar
elliptic-hyperbolic equation associated to electromagnetic wave
propagation in cold plasma, although a well-posed Dirichlet
problem for weak solutions has been formulated for data prescribed
only on part of the boundary.$^{24}$ (In fact, Magnanini and
Talenti do more than prove the existence of a weak solution: they
also show uniqueness and internal regularity modulo a point, and
give an explicit representation of the solution in terms of
special functions.)

The existence of a well-posed Dirichlet problem is important because
physical reasoning often suggests that the full Dirichlet problem is the
correct problem even in the case of equations for which mathematical
reasoning suggests otherwise.

Two questions suggested by Magnanini and Talenti's paper are:

\textit{i)} The transformation (18) itself fails at caustics
(which are not generally identical to the caustics of the physical
model). One would like to characterize regions at which this
linearization method fails and the nature of the singularities
that arise in such regions. See, for example, Proposition 2 of
Ref.\ 26.

\textit{ii)} The result proven in Ref.\ 18 requires the domain
boundary to lie entirely within the elliptic region of the
equation. It is an important quality of eq.\ (17) that the
elliptic region surrounds the hyperbolic region, a property not
shared by other elliptic-hyperbolic equations. Thus there is some
mathematical interest in asking whether solutions of (17) exist
with boundary points lying in both the elliptic and hyperbolic
regions, a situation in which this special condition is no longer
applicable. We consider this question in Sec. 3.3.

Equation (16) is a special case of the system

\begin{equation}
\left[ \left( p^{2}+q^{2}\right) ^{2}-q^{2}\right]
p_{x}+2pqp_{y}+\left[ \left( p^{2}+q^{2}\right) ^{2}-p^{2}\right]
q_{y}=0,
\end{equation}
\begin{equation}
p_{y}-q_{x}=0.
\end{equation}

This system is equivalent to eq.\ (16) if there is a continuously
differentiable scalar function $v\left( x,y\right) $ for which
$v_{x}=p$ and $v_{y}=q.$ (Such a function always exists locally,
by eq.\ (20).)

Consider any two-dimensional quasilinear system of two equations having the
form
\begin{equation}
\left[
\begin{array}{cc}
a_{11} & a_{12} \\
a_{21} & a_{22}
\end{array}
\right] \frac{\partial }{\partial x}\left(
\begin{array}{c}
p \\
q
\end{array}
\right) +\left[
\begin{array}{cc}
b_{11} & b_{12} \\
b_{21} & b_{22}
\end{array}
\right] \frac{\partial }{\partial y}\left(
\begin{array}{c}
p \\
q
\end{array}
\right) =\left(
\begin{array}{c}
0 \\
0
\end{array}
\right) ,
\end{equation}
where the entries of the coefficient matrices depend only on $p$
and $q.$ \ Then the coordinate transformation $\left( x,y\right)
\rightarrow \left( p,q\right)$ takes eq.\ (21) into the linear
form
\[
\left[
\begin{array}{cc}
b_{12} & -a_{12} \\
b_{22} & -a_{22}
\end{array}
\right] \frac{\partial }{\partial p}\left(
\begin{array}{c}
x \\
y
\end{array}
\right) +\left[
\begin{array}{cc}
-b_{11} & a_{11} \\
-b_{21} & a_{21}
\end{array}
\right] \frac{\partial }{\partial q}\left(
\begin{array}{c}
x \\
y
\end{array}
\right) =\left(
\begin{array}{c}
0 \\
0
\end{array}
\right) ,
\]
provided the Jacobian of the transformation
\[
J=\frac{\partial \left( x,y\right) }{\partial \left( p,q\right) }=\frac{%
\partial x}{\partial p}\frac{\partial y}{\partial q}-\frac{\partial y}{%
\partial p}\frac{\partial x}{\partial q}
\]
is nonzero. This special case of the Legendre transformation is
called a \textit{hodograph map}, and the space having coordinates
$(p,q)$ is called the \textit{hodograph plane;} see,
\textit{e.g.,} Sec.\ V.2.2 of Ref.\ 5.

The coordinate systems $(p,q)$ and $(x,y)$ are related by eq.\
(18),where
\[
\left( x,y\right) =\left( \frac{\partial V}{\partial p},\frac{\partial V}{%
\partial q}\right)
\]
and
\[
\left( p,q\right) =\left( \frac{\partial v}{\partial x},\frac{\partial v}{%
\partial y}\right) .
\]

Applying a hodograph transformation to eqs.\ (19), (20) yields the
system
\begin{equation}
\left[ \left( p^{2}+q^{2}\right) ^{2}-p^{2}\right] x_{p}-2pqx_{q}+\left[
\left( p^{2}+q^{2}\right) ^{2}-q^{2}\right] y_{q}=0,
\end{equation}
\begin{equation}
x_{q}-y_{p}=0.
\end{equation}
This system is equivalent to eq.\ (17) if there is a continuously
differentiable scalar function $V\left( x,y\right) $ for which
$V_{p}=x$ and $V_{q}=y.$ (Again, this can always be arranged
locally.)

As in Sec.\ 2, we write the second-order terms of eqs.\ (22), (23)
in the form $A^1 \mathbf{u}_{x}+A^2 \mathbf{u}_{y},$ where
$\mathbf{u}=\mathbf{u}(x,y)$ and in this case
\[
A^1=\left[
\begin{array}{cc}
\left( x^{2}+y^{2}\right) ^{2}-x^{2} & 0 \\
0 & -1
\end{array}
\right]
\]
and
\[
A^2=\left[
\begin{array}{cc}
-2xy & \left( x^{2}+y^{2}\right) ^{2}-y^{2} \\
1 & 0
\end{array}
\right] .
\]
The characteristic equation
\[
\left| A^1-\lambda A^2\right| =-\left\{ \left[ \left(
x^{2}+y^{2}\right) ^{2}-y^{2}\right] \lambda ^{2}+2xy\lambda
+\left[ \left( x^{2}+y^{2}\right) ^{2}-x^{2}\right] \right\}
\]
possesses two real roots $\lambda _{1},\lambda _{2}$ precisely when $%
x^{2}+y^{2}>\left( x^{2}+y^{2}\right) ^{2},$ that is, when $x^{2}+y^{2}<1.$
\ Thus the system is hyperbolic at points lying inside the open unit disc
centered at $\left( x,y\right) =\left( 0,0\right) $ and elliptic outside the
closure of this disc. The circle $x^{2}+y^{2}=1,$ along which the change in
type occurs, is the parabolic region of the system.

\subsection{Uniform asymptotic approximations}

Substitution of the simplest formula for an oscillatory wave into the wave
equation results in the \textit{Helmholtz equation}
\begin{equation}
\Delta U\left( \mathbf{x}\right) +k^{2}\nu ^{2}U\left( \mathbf{x}\right) =0,
\end{equation}
where we take $\mathbf{x}$ to be a vector in $\mathbb{R}^{2},$\
and where $k$ and $\nu $ are physical constants. In the standard
application, $\nu $ is the refractive index of the medium and $k$
is inversely proportional to wavelength. In the region of visible
light, the wavelength is sufficiently small that $k$ dominates
over all other mathematically relevant parameters, an undesirable
property known as \textit{stiffness.}

For this reason, short-wave solutions of (24) are usually
approximated by \textit{uniform asymptotic expansions}$^{12,17}$
which satisfy (24) to arbitrarily high order in $k^{-1}.$ These
approximations are valid in regions which contain smooth and
convex caustics such as a circular caustic. The size of the region
of validity is independent of $k.$ Take $\nu \equiv 1$ and
approximate the solution to (24) by an expansion having the form

\[
U_{approx}(x,y) =
\]
\[
\left\{ Z\left( k^{2/3}u\right) \left( \sum_{j=0}^{\infty }W_{j}\left(
\mathbf{r}\right) \cdot \left( ik\right) ^{-j}\right) +\frac{i}{k^{1/3}}%
Z^{\prime }\left( k^{2/3}u\right) \left( \sum_{j=0}^{\infty }X_{j}\left(
\mathbf{r}\right) \cdot \left( ik\right) ^{-j}\right) \right\}
\]
\[
\times \exp \left[ ikv\left( x,y\right) \right] ,
\]
where $u\left( x,y\right) ,$ $v\left( x,y\right) ,$ $W_{j}\left( \mathbf{r}%
\right) ,$ and $X_{j}\left( \mathbf{r}\right) $ are functions
which do not depend on $k$ and which are to be determined with the
solution; the function $Z(t)$ is a solution of the \textit{Airy
equation}
\[
Z^{\prime \prime }\left( t\right) - tZ\left( t\right) =0,
\]
with initial conditions
\[
Z(0)=\frac{3^{-2/3}}{\Gamma \left( 2/3\right) }
\]
and
\[
Z^{\prime }(0)=-\frac{3^{-1/3}}{\Gamma \left( 1/3\right) },
\]
where $\Gamma \left( \;\right) $ is the gamma function.

This model implies the following system of equations for $u$ and $v:$%
\[
u\left( u_{x}^{2}+u_{y}^{2}\right) -\left( v_{x}^{2}+v_{y}^{2}\right) +1=0,
\]
\[
u_{x}v_{x}+u_{y}v_{y}=0.
\]

In Ref.\ 18 three possible solutions of this system are
enumerated:
\[
u=0,\;\left| \nabla v\right| ^{2}=1;
\]
\[
\left| \nabla u\right| =0,\;\left| \nabla v\right| ^{2}=1;
\]
the third possibility is that eq.\ (16) is satisfied.

Obviously, the third alternative is the most interesting, and this
case is studied in Ref.\ 18. This case is linearized to eq.\ (17)
by a hodograph transformation.

\subsection{A first-order system}

Thus we are led to a system resembling eqs.\ (5)-(7):
\begin{equation}
L\mathbf{u}=\mathbf{g},
\end{equation}
where
\[
L=\left( L_{1},L_{2}\right) ,\;\mathbf{g}=\left( g_{1},g_{2}\right) ,
\]
\[
\mathbf{u}=\left( u_{1}\left( x,y\right) ,u_{2}\left( x,y\right)
\right) ,\;\left( x,y\right) \in \Omega \subset \subset
\mathbb{R}^{2},
\]
\begin{equation}
\left( L\mathbf{u}\right) _{1}=\left[ f\left( x,y\right) -x^{2}\right]
u_{1x}-2xyu_{1y}+\left[ f\left( x,y\right) -y^{2}\right] u_{2y}
\end{equation}
and
\begin{equation}
\left( L\mathbf{u}\right) _{2}=\left[ f\left( x,y\right) -y^{2}\right]
\left( u_{1y}-u_{2x}\right) ,
\end{equation}
for
\begin{equation}
f\left( x,y\right) =\left( x^{2}+y^{2}\right) ^{2}.
\end{equation}
The domain is chosen so that
\[
f\left( x,y\right) -y^{2}\ne 0,
\]
under which system (25)-(28) becomes an inhomogeneous
generalization of eqs.\ (22), (23). If in particular,
$g_{1}=g_{2}=0,$ $u_{1}=V_{x},$ and $u_{2}=V_{y}, $ where $V\left(
x,y\right) $ is a scalar function, then eqs.\ (25)-(28) reduce to
eq.\ (17).

As in the preceding sections, the second-order terms of eqs.\
(25)-(28) can be written in the form
$A^1\mathbf{u}_{x}+A^2\mathbf{u}_{y},$ where
\[
A^1=\left[
\begin{array}{cc}
f(x,y)-x^{2} & 0 \\
0 & -\left( f(x,y)-y^{2}\right)
\end{array}
\right]
\]
and
\[
A^2=\left[
\begin{array}{cc}
-2xy & f(x,y)-y^{2} \\
f(x,y)-y^{2} & 0
\end{array}
\right] .
\]
We find that the system is hyperbolic in the intersection of
$\Omega $ with the open unit disc centered at $\left( 0,0\right) $
and elliptic in the intersection of $\Omega $ with the complement
of the closure of this disc.

\subsection{Strong solutions in an annulus}

Writing eq.\ (17) in polar coordinates $(r,\theta),$ $r \geq 0,$
$0 < \theta \leq 2\pi,$ we obtain$^{18}$

\begin{equation}
    \left(r^2-1\right)V_{rr}+rV_r+V_{\theta\theta}=0.
\end{equation}
Letting $u_1 = V_r$ and $u_2 = V_\theta$ transforms eq.\ (29) into
a first-order system of the form

\begin{equation}
    L\mathbf{u}=A^1\mathbf{u}_r+A^2\mathbf{u}_\theta+B\mathbf{u}=\textbf{f},
\end{equation}
with $\textbf{u}=\left(u_1(r,\theta),u_2(r,\theta)\right),$
$\textbf{f}=(0,0),$
\begin{equation}
    A^1=\left(%
\begin{array}{cc}
  r^2-1 & 0 \\
  0 & -1 \\
\end{array}%
\right),\,A^2=\left(%
\begin{array}{cc}
  0 & 1 \\
  1 & 0 \\
\end{array}%
\right),
\end{equation}
and

\[
    B=\left(%
\begin{array}{cc}
  r & 0 \\
  0 & 0 \\
\end{array}%
\right).
\]

As in Sec.\ 2.2, the matrices are symmetric and we find that $Q=
2B^\ast-A_r^1-A_\theta^2$ is exactly zero, suggesting that an
arbitrarily small perturbation of the matrix $B$ will result in a
symmetric-positive operator. However, we find that we can retain
the consistency condition $u_{1\theta}-u_{2r}=0$ if we employ a
multiplier $E$ as described in Sec.\ 2.2. Thus we define

\[
E=\left(%
\begin{array}{cc}
  a & c\left(1-r^2\right) \\
  c & a \\
\end{array}%
\right),
\]
where $a=a(r,\theta)$ and $c=c(r,\theta)$ are continuously
differentiable functions to be chosen. We replace $B$ by the
matrix

\begin{equation}
    B_\varepsilon = \left(%
\begin{array}{cc}
  r+\varepsilon_1 & \varepsilon_2 \\
  0 & 0 \\
\end{array}%
\right),
\end{equation}
where $\varepsilon_1, \varepsilon_2$ are arbitrarily small,
positive constants.

Replacing eq.\ (30) by the system

\begin{equation}
    EL=EA^1\textbf{u}_r+EA^2\textbf{u}_\theta+EB_\varepsilon \textbf{u}=E\textbf{f},
\end{equation}
with $A^1,$ $A^2,$ and $B_\varepsilon$ given by eqs.\ (31) and
(32), we find that $EL$ is a symmetric-positive operator provided
we choose $0\leq\varepsilon_0\leq r \leq R < \infty,$ $c$ a
positive constant, and
\[
a=Me^{\varepsilon_2\theta}+\frac{\left(\sqrt{2}-\varepsilon_1\right)c}{\varepsilon_2},
\]
where $M$ is a constant such that $M>>c.$

We will solve eqs.\ (33) in the annulus $\Omega_5$ given by
$\varepsilon_0 \leq r \leq \sqrt{2}.$ (The solutions can be
patched into an elliptic boundary-value problem on the annulus
$\sqrt{2}\leq r\leq R.$) Data will be prescribed on the outer
boundary only. Annular domains are natural when numerical methods
are used to study an equation, such as eq.\ (17), which is known
to be singular at the origin, with the singular point excluded.
The problem is also of some historical interest. An equation
differing from (17) only in its lower-order terms was one of the
first elliptic-hyperbolic equations to be studied, more than 75
years ago, by Bateman (Sec.\ 9 of Ref.\ 1). That equation equation
arose from the solution of Laplace's equation in toroidal
coordinates.$^2$ At the time, Bateman raised the question of the
existence and uniqueness of solutions in an annular region
containing the unit circle, in which the outer boundary lies in
the elliptic region and the inner boundary lies in the hyperbolic
region of the equation.

Although the system that we consider is a small perturbation of
the one studied in Ref.\ 18, we note that the original equation is
itself an approximation, as described in Sec.\ 3.1.

\bigskip

\textbf{Theorem 3}. \emph{Equations (33) with boundary conditions}
\begin{equation}
    \tau(\theta)u_1+\sigma(\theta)u_2=0,\,\,\sigma^2(\theta)>\tau^2(\theta)
\end{equation}
\emph{imposed on the outer boundary $r=\sqrt{2},$ possess a strong
solution on the annulus $\Omega_5$ for every $\textbf{f} \in
L^2(\Omega_5).$}

\bigskip

\emph{Proof}. Although the equations are different, the argument
is similar to the proof by Torre$^{30}$ of the corresponding
assertion for the helically reduced wave equation.

The matrices $E$ and $B_\varepsilon$ have been constructed in such
a way that the operator $EL$ is manifestly symmetric-positive (for
large $M$), and the proof again reduces to a demonstration that
the boundary conditions are admissible. At the outer boundary,
choose $\textbf{n}_{outer}=dr.$ Then
\[
\beta_{outer}=\left(%
\begin{array}{cc}
  a & c \\
  c & -a \\
\end{array}%
\right).
\]
Choose
\[
\beta_{outer^-}=
\]
\[
\frac{1}{\sigma^2+\tau^2}\left(%
\begin{array}{cc}
  \sigma\tau c+\tau^2 a & \sigma^2c+\sigma\tau a \\
  -\sigma \tau a+\tau^2c & -\sigma^2 a+\sigma\tau c \\
\end{array}%
\right).
\]
Then
\[
\beta_{outer^+}=
\]
\[
\frac{1}{\sigma^2+\tau^2}\left(%
\begin{array}{cc}
  -\sigma\tau c+\sigma^2 a & \tau^2c-\sigma\tau a \\
  \sigma \tau a+\sigma^2c & -\tau^2 a-\sigma\tau c \\
\end{array}%
\right).
\]
Notice that $\beta_{outer^+}+\beta_{outer^-}=\beta_{outer}$ and
that $\beta_{outer^-}\textbf{u}=0,$ as (34) implies that
$u_2=-(\tau/\sigma)u_1$ on the circle $r=\sqrt{2}.$ Moreover,

\[
\mu =
\]
\[
\frac{1}{\sigma^2+\tau^2}\left(%
\begin{array}{cc}
  \left(\sigma^2-\tau^2\right)a-2\sigma\tau c & \left(\tau^2-\sigma^2\right)c-2\sigma\tau a \\
  \left(\sigma^2-\tau^2\right)c+2\sigma\tau a & \left(\sigma^2-\tau^2\right) a -2\sigma\tau c \\
\end{array}%
\right),
\]
implying that

\[
\mu^\ast =
\]
\[
\frac{1}{\sigma^2+\tau^2}\left(%
\begin{array}{cc}
  \left(\sigma^2-\tau^2\right)a-2\sigma\tau c & 0 \\
  0 & \left(\sigma^2-\tau^2\right) a -2\sigma\tau c \\
\end{array}%
\right).
\]
But this matrix is non-negative, given that $\sigma^2>\tau^2,$
provided that we choose $M$ sufficiently large.

On the inner boundary we choose
\[
\textbf{n}_{inner}=\left(\varepsilon_0^2-1\right)^{-1}dr.
\]
Then
\[
\beta_{inner}=\left(%
\begin{array}{cc}
  a & c \\
  c & -\left(\varepsilon_0^2-1\right)^{-1}a \\
\end{array}%
\right).
\]
Choose
\[
\beta_{inner^-} = \left(%
\begin{array}{cc}
  0 & 0 \\
  0 & 0 \\
\end{array}%
\right).
\]
Then $\beta_{inner^+}=\beta_{inner}$ and  $\mu^\ast\geq 0$ for $M$
sufficiently large. Moreover, $\beta_{inner^-} \textbf{u}=0$ on
the circle $r=\varepsilon_0$ for any vector $\textbf{u}.$

This completes the proof of Theorem 3.

\section{A remark on terminology and notation}

Hodge$^8$ originally considered a $p$-form $\omega $ to be
harmonic if it satisfies the first-order equations

\begin{equation}
d\omega =\delta \omega =0,
\end{equation}
where $d:\Lambda ^{p}\rightarrow \Lambda ^{p+1}$ is the exterior
derivative and $\delta :\Lambda ^{p+1}\rightarrow \Lambda ^{p}$ is
the adjoint of $d.$ \ If the underlying space is $\mathbb{R}^{2}$
and $\omega $ is a 1-form given by
\[
\omega =pdx+qdy,
\]
where $p$ and $q$ are continuously differentiable functions, then
the Hodge equations (35) reduce to the Cauchy-Riemann equations
for $p$ and $-q.$ However, although $d$ is independent of the
underlying metric, its adjoint $\delta $ has a different local
form for different metrics. Thus for a surface having metric
tensor $g_{ij}$, the Hodge equations for 1-forms are equivalent to
the system (3), (4). A discussion of exterior forms and their
properties is given in, \emph{e.g.,} Ref.\ 21.

The standard definition of a \textit{harmonic form} is given in
terms of a second-order operator: it is a solution of the
form-valued Laplace-Beltrami equations
\[
\left( d\delta +\delta d\right) \omega =0.
\]
If the domain has \textit{zero boundary} (either no boundary or
the prescribed value $\omega \equiv 0$ on the boundary), \ then
the definitions in terms of first- and second-order operators are
equivalent. Otherwise, one distinguishes them by calling a form
that satisfies eqs.\ (35) a \textit{harmonic field}. In words, the
Hodge equations assert that a harmonic field $\omega $ is both
\emph{closed} ($d\omega =0$) and \emph{co-closed} ($\delta \omega
=0$) under the exterior derivative $d.$ Obviously, every harmonic
field is a harmonic form, but the converse is false.

Notice that in eqs.\ (6) and (7), $L_1 \ne \delta$ and $L_2 \ne
d.$ For example, $L_2$ includes a factor of $1-y^2$ whereas $d$
does not, and $\delta$ includes determinants of the metric tensor,
whereas $L_1$ does not. In addition, cases 1 and 2 of (6), (7)
include additional lower-order terms. Thus for example $\delta$
and $d$ are self-adjoint, whereas $L_1$ and $L_2$ are not unless
$k_1=k_2=k_3=k_4=0.$

\bigskip

\textbf{Acknowledgment}. I am grateful to an anonymous referee for
helpful criticism of an earlier draft of this paper.

\section{References}

$^1$Bateman, H., ``Notes on a differential equation which occurs in
the two-dimensional motion of a compressible fluid and the
associated variational problems,'' \emph{Proc.\ R.\ Soc.\ London
Ser.} \textbf{A}, \textbf{125}, 598-618 (1929).

\bigskip

\noindent $^2$Bateman, H. \emph{Partial Differential Equations},
Dover, New York, 1944.

\bigskip

\noindent $^3$Beltrami, E., ``Saggio di interpretazione della
geometria non-euclidea,'' \emph{Giornale di Matematiche} \textbf{6},
284-312, 1868.

\bigskip

\noindent $^4$Bers, L., \emph{Mathematical Aspects of Subsonic and
Transonic Gas Dynamics,} Wiley, New York, 1958.

\bigskip

\noindent $^5$Courant, R. and D. Hilbert, \emph{Methods of
Mathematical Physics, Vol. 2,} Wiley-Interscience, New York, 1962.

\bigskip

\noindent $^6$Dencker, N., ``On the propagation of polarization sets
for systems of real principal type,'' \emph{J.\ Functional Analysis}
\textbf{46}, 351-372 (1982).

\bigskip

\noindent $^7$Friedrichs, K. O., ``Symmetric positive linear
differential equations,'' \emph{Commun.\ Pure Appl.\ Math.}
\textbf{11} 333-418 (1958).

\bigskip

\noindent $^8$Hodge, W. V. D., ``A Dirichlet problem for harmonic
functionals with applications to analytic varieties,'' \emph{Proc.\
London Math.\ Soc.} \textbf{36}, 257-303 (1934).

\bigskip

\noindent $^9$Hua, L. K., ``Geometrical theory of partial
differential equations,'' in: \emph{Proceedings of the 1980 Beijing
Symposium on Differential Geometry and Differential Equations} (S.
S. Chern and Wu Wen-ts\"{u}n, eds.), Gordon and Breach, New York,
1982, pp.\ 627-654.

\bigskip

\noindent $^{10}$Ji, X-H., and D-Q.Chen, ``Tricomi's problems of
non-homogeneous equation of mixed type in real projective plane,''
in: \emph{Proceedings of the 1980 Beijing Symposium on Differential
Geometry and Differential Equations} (S. S. Chern and Wu
Wen-ts\"{u}n, eds.), Gordon and Breach, New York, 1982, pp.\
1257-1271.

\bigskip

\noindent $^{11}$Ji, X-H., and D-Q. Chen, ``The Tricomi's problem of
the nonhomogeneous equation of mixed type in the real projective
plane,'' in: \emph{Mixed Type Equations} (J. M. Rassias, ed.),
Teubner, Leipzig, 1986, pp.\ 280-300.

\bigskip

\noindent $^{12}$Kravtsov, Yu. A., ``A modification of the
geometrical optics method'' [in Russian], \textit{Radiofizika}
\textbf{7} 664-673 (1964).

\bigskip

\noindent $^{13}$Kravtsov, Yu. A. and Yu. I. Orlov,
\textit{Caustics, Catastrophes, and Wave Fields}, Springer-Verlag,
New York, 1999.

\bigskip

\noindent $^{14}$Ladyzhenskaya, O. A., and N.N. Ural'tseva,
\emph{Linear and Quasilinear Elliptic Equations}, Academic Press,
New York, 1968.

\bigskip

\noindent $^{15}$Lax, P. D., and R. S. Phillips, ``Local boundary
conditions for dissipative symmetric linear differential
operators,'' \emph{Commun.\ Pure Appl.\ Math.} \textbf{13}, 427-455
(1960).

\bigskip

\noindent $^{16}$Lin, C. S., ``The local isometric embedding in
$R^3$ of two-dimensional Riemannian manifolds with Gaussian
curvature changing sign cleanly,'' \emph{Commun.\ Pure Appl.\ Math.}
\textbf{39} 867-887 (1986).

\bigskip

\noindent $^{17}$Ludwig, D., ``Uniform asymptotic expansions at a
caustic,'' \textit{Commun.\ Pure Appl.\ Math.} \textbf{19}, No. 2,
215-250 (1966).

\bigskip

\noindent $^{18}$Magnanini, R., and G. Talenti, ``Approaching a
partial differential equation of mixed elliptic-hyperbolic type,''
in \textit{Ill-posed and Inverse Problems}, S. I. Kabanikhin and V.
G. Romanov, eds., VSP 2002, pp.\ 263-276.

\bigskip

\noindent $^{19}$Morawetz, C. S., ``A weak solution for a system of
equations of elliptic-hyperbolic type,'' \emph{Commun.\ Pure Appl.\
Math.} \textbf{11}, 315-331 (1958).

\bigskip

\noindent $^{20}$Morawetz, C. S., ``The Dirichlet problem for the
Tricomi equation,'' \emph{Commun.\ Pure Appl.\ Math.} \textbf{23},
587-601 (1970).

\bigskip

\noindent $^{21}$Morita, S., \emph{Geometry of Differential
Forms}, American Mathematical Society, Providence, 2001.

\bigskip

\noindent $^{22}$Nye, J. F., \textit{Natural Focusing and the Fine
Structure of Light,} Institute of Physics Publishing, Bristol,
1999.

\bigskip

\noindent $^{23}$Otway, T. H., ``Hodge equations with change of
type,'' \emph{Annali Mat.\ Pura ed Applicata} \textbf{181}, 437-452
(2002).

\bigskip

\noindent $^{24}$Otway, T. H., ``A boundary-value problem for cold
plasma dynamics,'' \emph{J.\ Appl.\ Math.} \textbf{3}, 17-33 (2003).

\bigskip

\noindent $^{25}$Otway, T. H., ``Maps and fields with compressible
density,'' \emph{Rendiconti Sem.\ Mat.\ Univ.\ Padova} \textbf{111},
133-159 (2004).

\bigskip

\noindent $^{26}$Otway, T. H., ``Geometric analysis near and across
a sonic curve,'' in: \emph{New Developments in Mathematical Physics
Research} (C. V. Benton, ed.), Nova Science Publishers, New York,
2004, pp.\ 27-54.

\bigskip

\noindent $^{27}$Payne, K. R., ``Interior regularity of the
Dirichlet problem for the Tricomi equation,'' \emph{J.\ Math.\
Anal.\ Appl.} \textbf{199}, 271-292 (1996).
\bigskip

\noindent $^{28}$Stillwell, J., \emph{Geometry of Surfaces},
Springer, New York, 1992.

\bigskip

\noindent $^{29}$Stoker, J. J., \textit{Water Waves},
Interscience, New York, 1957.

\bigskip

\noindent $^{30}$Torre, C. G., ``The helically-reduced wave
equation as a symmetric-positive system,'' \emph{J. Math. Phys.}
\textbf{44}, 6223-6232 (2003).

\bigskip

\noindent $^{31}$Tricomi, F., ``Sulle equazioni lineari alle
derivate parziali di secondo ordine, di tipo misto,''
\emph{Rendiconti Atti dell' Accademia Nazionale dei Lincei,} Ser.\
5, \textbf{14}, 134-247 (1923).

\bigskip

\noindent $^{32}$Veblen, O, and J. W. Young, \emph{Projective
Geometry}, Ginn and Co., Boston, 1918.

\end{document}